\documentclass[twocolumn,pra,showpacs,superscriptaddress]{revtex4-1}
\usepackage{amssymb}
\usepackage{amsmath}
\usepackage{graphicx}
\usepackage{epsfig}

\begin{document}

\title{Equivalent spin-orbit interaction in two-polariton
Jaynes-Cummings-Hubbard model}
\author{C. Li}
\affiliation{School of Physics, Nankai University, Tianjin 300071, China}
\author{X. Z. Zhang}
\affiliation{College of Physics and Materials Science, Tianjin Normal University, Tianjin
300387, China}
\author{Z. Song}
\email{songtc@nankai.edu.cn}
\affiliation{School of Physics, Nankai University, Tianjin 300071, China}

\begin{abstract}
A hybrid quantum system combines two or more distinct quantum components,
exhibiting features not seen in these individual systems. In this work, we
study the one-dimensional Jaynes-Cummings-Hubbard model in the
two-excitation subspace. We find that the center momentum of two-excitation
induces a magnetic flux piercing the $4$-leg ladder in the auxiliary space.
Furthermore, it is shown that the system in $\pi $-center-momentum subspace
is equivalent to a chain system for spin-$1$\ particle with spin-orbit
coupling. As a simple application, based on this concise description, a
series of bound-pair eigenstates is presented, which displays long-range
correlation.
\end{abstract}

\maketitle


\section{Introduction}

A hybrid quantum system combines two or more distinct quantum components,
exhibiting features not seen in these individual systems. This provides a
promising platform to study novel quantum phenomena. The
Jaynes-Cummings-Hubbard (JCH) model is an archetype of such hybridization,
which consists of the JCH and the coupled cavities. This model was proposed
for the use of the atom-light interaction in coupled microcavity arrays to
create strongly correlated many-body models \cite{Hartmann, Angelakis,
greentree}. It has received intensive study in a variety of directions (See
the review \cite{Hartmann1, Tomadin} and references therein).

The researches mainly focus on the ground state phase of many-particle
system and the dynamics in single-particle system. Mott insulator phase and
superfluid phase are identified\ by the traditional order parameter, the
average of the annihilation operator \cite{greentree} and the observable
quantities, atomic concurrence and photon visibility \cite{Huo}. The
single-particle dynamics suggests that this hybrid architecture can be the
quantum coherent device to transfer and store quantum information as well as
to create the laser-like output \cite{zhou1, zhou2, hu}. Recently, few-body
problem for the JCH Hamiltonian is also investigated \cite{Max, Peng},
claiming the existence of two-polariton bound states when the photon-atom
interaction is sufficiently strong.

In this work, we study the one-dimensional JCH model in the two-excitation
subspace. In each invariant subspace, the sub-Hamiltonian is equivalent to a
$4$-leg ladder with an effective flux, which is proportional to the center
momentum of two excitations. It is shown that in $\pi $-center-momentum
subspace, the ladder system can be reduced to a chain system of spin-$1$\
particle with spin-orbit coupling. As a simple application, based on this
concise description, a series of bound-pair eigenstates is presented, which
display long-range polaritonic entanglement.

This paper is organized as follows. In Section \ref{JC-Hubbard model}, we
present the JCH\ model and the basis set. In Section \ref{4-leg ladder with
flux}, the equivalent Hamiltonian is given. In Section \ref{Equivalent
Hamiltonian}, in $\pi $-center-momentum subspace, the equivalent Hamiltonian
is reduced to a chain system of spin-$1$\ particle with spin-orbit coupling.
In Section \ref{Exact bound-pair states}, a series of bound-pair eigenstates
is constructed. In Section \ref{Long-range entanglement}, we investigate the
quantum correlation for the bound-pair states. Finally, we give a summary
and discussion in Section \ref{Summary}.

\section{JCH model}

\label{JC-Hubbard model}

The Jaynes--Cummings model describes a cavity array doped with a single
two-level atoms embedded in each cavity and the dipole interaction leads to
dynamics involving photonic and atomic degrees of freedom, which is in
contrast to the widely studied Bose-Hubbard model. Such hybrid system can be
implemented with the defect array in photonic crystal \cite{fan} or
Josephson junction array in cavity \cite{zhou1}. The Hamiltonian of a hybrid
system, or a lattice atom-photon system,
\begin{equation}
H=H_{\mathrm{0}}+H_{\mathrm{JC}}+H_{\mathrm{C}}  \label{H}
\end{equation}%
can be written as three parts, free Hamiltonians of atom and photon,%
\begin{equation}
H_{\mathrm{AP}}=\omega _{a}\sum_{l=1}^{N}a_{i}^{\dag }a_{i}+\omega
_{b}\sum_{l=1}^{N}\left\vert e\right\rangle _{l}\left\langle e\right\vert ,
\end{equation}%
the JC type cavity-atom interaction in the $i$-th defect
\begin{equation}
H_{\mathrm{JC}}=\lambda \sum_{l=1}^{N}\left( a_{l}^{\dag }\left\vert
g\right\rangle _{l}\left\langle e\right\vert +\text{\textrm{H.c.}}\right) ,
\label{H2}
\end{equation}%
with strength $\lambda $ and the photon hopping between nearest neighbor
cavities
\begin{equation}
H_{\mathrm{C}}=-\kappa \sum_{l=1}^{N}\left( a_{l}^{\dag }a_{l+1}+\text{%
\textrm{H.c.}}\right) ,
\end{equation}%
with hopping integral constant $\kappa $ for the tunneling between adjacent
cavities. Here, $\left\vert g\right\rangle _{i}$ ($\left\vert e\right\rangle
_{i}$) denotes the ground (excited) state of the atom placed at $i$th
cavity; $a_{i}^{\dag }$ and$\ a_{i}$ are the creation and annihilation
operators of a photon at defect $i$. Obviously the total excitation number
\begin{equation}
\mathcal{\hat{N}}\mathcal{=}\sum_{i=1}\mathcal{\hat{N}}_{i}=\sum_{i=1}\left(
a_{i}^{\dag }a_{i}+\sigma _{i}^{z}+\frac{1}{2}\right) ,
\end{equation}%
is conserved quantity for the Hamiltonian $H$, i.e., $[H,\mathcal{\hat{N}}%
]=0 $, where $\sigma _{l}^{z}\left\vert e\right\rangle _{l}=\left\vert
e\right\rangle _{l}$ and $\sigma _{l}^{z}\left\vert g\right\rangle
_{l}=-\left\vert g\right\rangle _{l}$. It can be seen that $\mathcal{\hat{N}}
$ is just the single excitation number of the polaritons. For each cavity,
the basis state can be expressed as $\left\{ \left\vert n\right\rangle
_{l}\left\vert e\right\rangle _{l},\left\vert n\right\rangle _{l}\left\vert
g\right\rangle _{l}\right\} $, where the basis state of the Fock space for $%
l $-th cavity is $\left\vert n\right\rangle _{l}=\left( a_{l}^{\dag }\right)
^{n}/\sqrt{n!}\left\vert 0\right\rangle _{l}$. In this paper, we consider
the invariant subspace with $\mathcal{N}=2$, which is spanned by the basis
in the form
\begin{equation}
\left(
\begin{array}{c}
\left\vert 2\right\rangle _{i} \\
\left\vert 1\right\rangle _{i}\left\vert 1\right\rangle _{i+j} \\
\left\vert e\right\rangle _{i}\left\vert e\right\rangle _{i+j} \\
\left\vert e\right\rangle _{i}\left\vert 1\right\rangle _{i^{\prime }}%
\end{array}%
\right) \equiv \left(
\begin{array}{c}
\left\vert 2\right\rangle _{i}\left\langle 0\right\vert \\
\left\vert 1\right\rangle _{i}\left\vert 1\right\rangle _{i+j}\left\langle
0\right\vert _{i}\left\langle 0\right\vert \\
\left\vert e\right\rangle _{i}\left\vert e\right\rangle _{i+j}\left\langle
g\right\vert _{i}\left\langle g\right\vert \\
\left\vert e\right\rangle _{i}\left\vert 1\right\rangle _{i^{\prime
}}\left\langle 0\right\vert _{i}\left\langle g\right\vert%
\end{array}%
\right) \left\vert G\right\rangle ,  \label{basis}
\end{equation}%
where $j\geqslant 1$ and $\left\vert G\right\rangle \equiv
\prod\nolimits_{i=1}\left\vert g\right\rangle _{i}\left\vert 0\right\rangle
_{i}$ denotes the empty state with zero $\mathcal{N}$. We denote the matrix
representation of the Hamiltonian of Eq. (\ref{H}) in the basis of Eq. (\ref%
{basis}) as $\underline{H}$. In the case of real values of $\kappa $\ and $%
\lambda $, we have $\underline{H}^{\ast }=\underline{H}$, which indicates
that $\underline{H}$\ has time-reversal symmetry.

\section{4-leg ladder with flux}

\label{4-leg ladder with flux}

\begin{figure}[tbp]
\includegraphics[ bb=19 84 561 779, width=0.5\textwidth, clip]{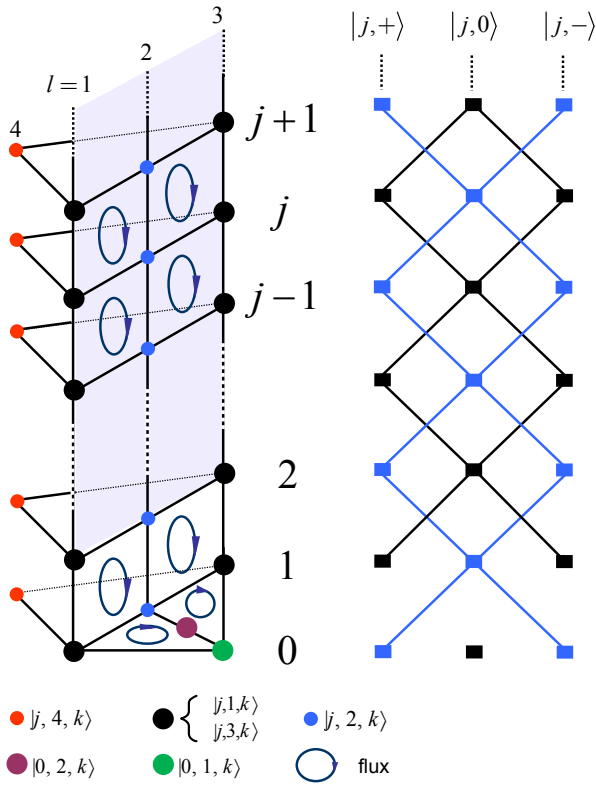}
\caption{(Color online) Schematic of the structures of equivalent
Hamiltonians for the one-dimensional JC-Hubbard model with two polaritons.
(a) In the invariant subspace with center momentum $k$, the equivalent
Hamiltonian $H_{k}$ describes a four-leg ladder with $k$-dependent flux. The
shadow indicates the semi-infinite uniform ladder. (b) For $k=\protect\pi $,
it is equivalent to a spin-$1$ chain with spin-orbit interaction. The graph
of $H_{\protect\pi }$\ consists of two unconnected subgraphs, characterized
by the parity $\Pi =\pm 1$. It indicates that $H_{\protect\pi }$\ can be
further decomposed into two independent parts $H_{\mathrm{o}}$\ (blue) and $H_{\mathrm{e}}$ (dark).}
\label{fig1}
\end{figure}

The system in translational invariant \cite{JLBP}. In the two-particle
Hilbert space, the Hamiltonian $H$ can be written as $H=\sum_{k}H_{k}$, where

\begin{eqnarray}
H_{k} &=&\sum_{j=1}\sum_{l=1}^{4}(J_{l}\left\vert j,l,k\right\rangle
\left\langle j+1,l,k\right\vert +\lambda \left\vert j,l,k\right\rangle
\left\langle j,l+1,k\right\vert  \notag \\
&&+\mathrm{H.c.})+\sum_{j=1}\sum_{l=1}^{4}(\mu _{l}\left\vert
j,l,k\right\rangle \left\langle j,l,k\right\vert )+h_{k},  \label{H_k}
\end{eqnarray}%
and
\begin{eqnarray}
h_{k} &=&\sum_{j=0,\text{ }l=1,3}J_{l}\left\vert j,l,k\right\rangle
\left\langle j+1,l,k\right\vert +  \notag \\
&&\sqrt{2}\lambda \left\vert 0,1,k\right\rangle \left\langle
0,2,k\right\vert +\sqrt{2}J_{2}\left\vert 0,2,k\right\rangle \left\langle
1,2,k\right\vert +  \label{h_end} \\
&&\sum_{j=0,\text{ }l=1,2}\mu _{l}\left\vert j,l,k\right\rangle \left\langle
j,l,k\right\vert +\mathrm{H.c.}  \notag
\end{eqnarray}%
where we have taken $\left\vert j,5,k\right\rangle \equiv \left\vert
j,1,k\right\rangle $ for $j\geq 1$, and $\left\vert 0,1,k\right\rangle
\equiv \left\vert 0,3,k\right\rangle $. The parameters reads

\begin{equation}
J_{1,2,3,4}=\left( -\kappa e^{ik/2},-2\kappa \cos \left( k/2\right) ,-\kappa
e^{-ik/2},0\right) ,
\end{equation}%
and%
\begin{equation}
\mu _{1,2,3,4}=\left( \omega _{a}+\omega _{b},2\omega _{a},\omega
_{a}+\omega _{b},2\omega _{b}\right) .
\end{equation}%
Here the set of states $\left\{ \left\vert j,l,k\right\rangle \right\} $\ is
defined as following: For $j\geq 1$, it reads%
\begin{equation}
\left(
\begin{array}{c}
\left\vert j,1,k\right\rangle \\
\left\vert j,2,k\right\rangle \\
\left\vert j,3,k\right\rangle \\
\left\vert j,4,k\right\rangle%
\end{array}%
\right) =\sum_{l}\frac{e^{ik\left( l+j/2\right) }}{\sqrt{N}}\left(
\begin{array}{c}
\left\vert 1\right\rangle _{l}\left\vert e\right\rangle _{l+j} \\
\left\vert 1\right\rangle _{l}\left\vert 1\right\rangle _{l+j} \\
\left\vert e\right\rangle _{l}\left\vert 1\right\rangle _{l+j} \\
\left\vert e\right\rangle _{l}\left\vert e\right\rangle _{l+j}%
\end{array}%
\right) ,  \label{basis_k1}
\end{equation}%
and%
\begin{equation}
\left(
\begin{array}{c}
\left\vert 0,1,k\right\rangle \\
\left\vert 0,2,k\right\rangle%
\end{array}%
\right) =\sum_{l}\frac{e^{ikl}}{\sqrt{N}}\left(
\begin{array}{c}
\left\vert 1\right\rangle _{l}\left\vert e\right\rangle _{l} \\
\left\vert 2\right\rangle _{l}%
\end{array}%
\right) ,  \label{basis_k2}
\end{equation}%
The expression of $H_{k}$ in Eq. (\ref{H_k}) has a clear physical meaning: $%
\left\vert j,l,k\right\rangle $ denotes the site state for $j$th site on the
$l$ leg of a $4$-leg ladder system with the effective magnetic piercing the
plaquette. The flux is proportional to the center momentum of two
excitations. The structure of $H_{k}$\ is schematically illustrated in Fig. %
\ref{fig1}. We note that the matrix representation of $H_{k}$ in the basis
of Eqs. (\ref{basis_k1}) and (\ref{basis_k2}), $\underline{H_{k}}$ breaks
the time-reversal symmetry. Nevertheless, we still have $\sum_{k}\underline{%
H_{k}}=\sum_{k}\underline{H_{k}}^{\ast }$ due to the fact that $\underline{%
H_{k}}^{\ast }$ $=\underline{H_{-k}}$ $=\underline{H_{4\pi -k}}$. In
essence, the nonzero plaquette flux arises from the relation between the
complex coupling constants $J_{1}=J_{3}^{\ast }=-\kappa e^{ik/2}$. In
contrast, one can see from $H_{k}$\ that the complex $\lambda $\ cannot
induce a nonzero plaquette flux. We would like to point that the effective
magnetic field in the present model is intrinsic, different from that
discussed in Refs. \cite{Jaeyoon} and \cite{Andrew}.

In order to understand the mechanism of the effective flux, we investigate
the exchange process between photon and atomic excitations from the state $%
\left\vert \psi \left( l,l+j\right) \right\rangle $ to state $\left\vert
\psi \left( l+j,l\right) \right\rangle $, where
\begin{equation}
\left\vert \psi \left( l,l^{\prime }\right) \right\rangle =(\left\vert
1\right\rangle _{l}\left\vert e\right\rangle _{l^{\prime }}-\left\vert
1\right\rangle _{l+1}\left\vert e\right\rangle _{l^{\prime }+1})/\sqrt{2}.
\end{equation}%
The action of $H$ provides at least two paths for this task: The first one
is described as%
\begin{eqnarray}
\text{I}\text{: } &&\left\vert \psi \left( l,l+j\right) \right\rangle \\
&\rightarrow &(\left\vert 1\right\rangle _{l}\left\vert 1\right\rangle
_{l+j}-\left\vert 1\right\rangle _{l+1}\left\vert 1\right\rangle _{l+j+1})/%
\sqrt{2}  \notag \\
&\rightarrow &\left\vert \psi \left( l+j,l\right) \right\rangle ,  \notag
\end{eqnarray}%
and another one is%
\begin{eqnarray}
\text{II}\text{: } &&\left\vert \psi \left( l,l+j\right) \right\rangle \\
&\rightarrow &(\left\vert 1\right\rangle _{l+1}\left\vert e\right\rangle
_{l+j}-\left\vert 1\right\rangle _{l}\left\vert e\right\rangle _{l+j+1})/%
\sqrt{2}  \notag \\
&\rightarrow &(\left\vert 1\right\rangle _{l+1}\left\vert 1\right\rangle
_{l+j}-\left\vert 1\right\rangle _{l}\left\vert 1\right\rangle _{l+j+1})/%
\sqrt{2}  \notag \\
&\rightarrow &(\left\vert 1\right\rangle _{l+1}\left\vert 1\right\rangle
_{l+j+1}-\left\vert 1\right\rangle _{l}\left\vert 1\right\rangle _{l+j})/%
\sqrt{2}  \notag \\
&\rightarrow &e^{i\pi }\left\vert \psi \left( l+j,l\right) \right\rangle .
\notag
\end{eqnarray}%
It shows that the exchange process acquires a phase $\pi $\ along the path
II, which is equivalent to the effect of a flux piercing the loop of two
paths. This investigation implies that the origin of the effective magnetic
field may be the special statistics property of the atomic excitations: acts
as a fermion at the same site but boson for different locations.

Based on the above analysis, the two-polariton scattering problem can be
reduced to the study of the single-particle time evolution in the four-leg
ladder system. In this paper, we only consider the eigen problem of the
two-polariton JC-Hubbard model.

\section{Equivalent Hamiltonian in $\protect\pi $-momentum subspace}

\label{Equivalent Hamiltonian}

We focus on the case of $k=\pi $ and $\omega _{a}=\omega _{b}$, which leads
to $H_{\mathrm{AP}}=\omega _{a}\mathcal{\hat{N}}$. It is a simple but
non-trivial case, since the hopping along the leg $2$ is switched off but
the plaquette flux still takes effect.\ We note that the on-site potentials $%
\mu _{l}$ of different legs are identical, which allows us to ignore the
diagonal terms in $H_{\pi }$.

Introducing the $3$-D vector bra and ket for
\begin{eqnarray}
\underline{\left\vert j\right\rangle } &=&\left(
\begin{array}{ccc}
\left\vert j,+\right\rangle , & \left\vert j,0\right\rangle , & \left\vert
j,-\right\rangle%
\end{array}%
\right) , \\
\underline{\left\langle j\right\vert } &=&\left(
\begin{array}{c}
\left\langle j,+\right\vert \\
\left\langle j,0\right\vert \\
\left\langle j,-\right\vert%
\end{array}%
\right) ,  \notag
\end{eqnarray}%
the Hamiltonian $H_{\pi }$ in the $\pi $-momentum subspace can be expressed
as%
\begin{equation}
H_{\pi }=H_{\mathrm{SO}}+0\sum_{j=1}\left\vert \psi _{j}\right\rangle
\left\langle \psi _{j}\right\vert ,  \label{H_pi}
\end{equation}%
with $[H_{\mathrm{SO}},\sum_{j=1}\left\vert \psi _{j}\right\rangle
\left\langle \psi _{j}\right\vert ]=0$, which indicates that $H_{\pi }$\ is
block-diagonal. The sub-Hamiltonian $H_{\mathrm{SO}}$\ is in the form
\begin{eqnarray}
H_{\mathrm{SO}} &=&\sqrt{2}i\kappa \underline{\left\vert 0\right\rangle }%
S_{x}\left( 1-S_{z}^{2}\right) \underline{\left\langle 1\right\vert }
\label{H_SO} \\
&&i\kappa \sum_{j=1}\underline{\left\vert j\right\rangle }S_{x}\underline{%
\left\langle j+1\right\vert }+\mathrm{H.c.}  \notag \\
&&+\sqrt{2}\lambda \underline{\left\vert 0\right\rangle }S_{z}\underline{%
\left\langle 0\right\vert }+2\lambda \sum_{j=1}\underline{\left\vert
j\right\rangle }S_{z}\underline{\left\langle j\right\vert },  \notag
\end{eqnarray}%
where the Pauli spin matrices for a spin-$1$ particle are given by%
\begin{equation}
S_{x,z}=\frac{1}{\sqrt{2}}\left(
\begin{array}{ccc}
0 & 1 & 0 \\
1 & 0 & 1 \\
0 & 1 & 0%
\end{array}%
\right) \text{, }\left(
\begin{array}{ccc}
1 & 0 & 0 \\
0 & 0 & 0 \\
0 & 0 & -1%
\end{array}%
\right) .
\end{equation}%
\ \ Here $\left\vert j,S_{z}\right\rangle $\ represents spin-$1$ particle at
$j$th site with spin polarization $S_{z}=0,\pm 1$ and is defined as%
\begin{equation}
\left\{
\begin{array}{c}
\left\vert j,\pm \right\rangle =(\left\vert j,1,\pi \right\rangle
+\left\vert j,3,\pi \right\rangle \\
\pm \left\vert j,2,\pi \right\rangle \pm \left\vert j,4,\pi \right\rangle
)/2, \\
\left\vert j,0\right\rangle =\left( \left\vert j,1,\pi \right\rangle
-\left\vert j,3,\pi \right\rangle \right) /\sqrt{2},%
\end{array}%
\right.
\end{equation}%
for $j\geq 1$, and%
\begin{equation}
\left\vert 0,\pm \right\rangle =\left( \left\vert 0,1,\pi \right\rangle \pm
\left\vert 0,2,\pi \right\rangle \right) /\sqrt{2}.
\end{equation}%
In addition, state $\left\vert \psi _{j}\right\rangle $\ $\left( j\geq
1\right) $\ is defined as%
\begin{eqnarray}
\left\vert \psi _{j}\right\rangle &=&\frac{1}{\sqrt{2}}(\left\vert j,2,\pi
\right\rangle -\left\vert j,4,\pi \right\rangle )  \label{Psi_j} \\
&=&\sum_{l}\frac{e^{i\pi l}}{\sqrt{2N}}\left( \left\vert 1\right\rangle
_{l}\left\vert 1\right\rangle _{l+j}-\left\vert e\right\rangle
_{l}\left\vert e\right\rangle _{l+j}\right) ,  \notag
\end{eqnarray}%
which constructs the complete orthogonal set together with states $\left\{
\left\vert j,\pm \right\rangle ,\left\vert j,0\right\rangle \right\} $. Of
particular interest, $\left\vert \psi _{j}\right\rangle $\ is the eigenstate
of $H$ with energy $2\omega _{a}$. In the expression of Eq. (\ref{H_pi}),
the zero-energy term represents this point, where we have ignored a constant
shift $2\omega _{a}$. We will discuss this problem in next section.

Consequently, within a specific invariant subspace, the system made of $N$\
cavity array with a single two-level atoms embedded in each cavity\textbf{\ }%
appears to be equivalent to a tight-binding chain for spin-$1$ particle with
spin-orbit interaction. The structures of $H_{\mathrm{SO}}$ is schematically
illustrated in Fig. \ref{fig1}. Intuitively, the graph of $H_{\mathrm{SO}}$\
consists of two unconnected subgraphs. This can be seen by observing that
the parity operator%
\begin{equation}
\widehat{\Pi }=\left( -1\right) ^{j+S_{z}+1}
\end{equation}%
with $\widehat{\Pi }\left\vert j,S_{z}\right\rangle =\Pi \left\vert
j,S_{z}\right\rangle $\ and $\Pi =\pm 1$, characterizing the two subgraphs.

Then we conclude that the equivalent Hamiltonian $H_{\mathrm{SO}}$ can be
decomposed into two independent parts\
\begin{equation}
H_{\mathrm{SO}}=H_{\mathrm{o}}+H_{\mathrm{e}},
\end{equation}%
with $\left[ H_{\mathrm{o}},H_{\mathrm{e}}\right] =0$, and $[\widehat{\Pi }%
,H_{\mathrm{e}}]=$ $[\widehat{\Pi },H_{\mathrm{o}}]=0$. The sub-Hamiltonians
are defined as%
\begin{eqnarray}
H_{\mathrm{o}} &=&i\kappa \sum_{j=1,3,5,\ldots }\underline{\left\vert
j\right\rangle }S_{x}\left( 1-S_{z}^{2}\right) \underline{\left\langle
j+1\right\vert }  \label{H_o} \\
&&i\kappa \sum_{j=2,4,6,\ldots }\underline{\left\vert j\right\rangle }%
S_{x}S_{z}^{2}\underline{\left\langle j+1\right\vert }+\mathrm{H.c.}  \notag
\\
&&+2\lambda \sum_{j=1,3,\ldots }\underline{\left\vert j\right\rangle }S_{z}%
\underline{\left\langle j\right\vert },  \notag
\end{eqnarray}%
and%
\begin{eqnarray}
H_{\mathrm{e}} &=&\sqrt{2}i\kappa \underline{\left\vert 0\right\rangle }%
S_{x}\left( 1-S_{z}^{2}\right) \underline{\left\langle 1\right\vert }
\label{H_e} \\
&&i\kappa \sum_{j=2,4,6,\ldots }\underline{\left\vert j\right\rangle }%
S_{x}\left( 1-S_{z}^{2}\right) \underline{\left\langle j+1\right\vert }
\notag \\
&&i\kappa \sum_{j=1,3,5,\ldots }\underline{\left\vert j\right\rangle }%
S_{x}S_{z}^{2}\underline{\left\langle j+1\right\vert }+\mathrm{H.c.}  \notag
\\
&&+\sqrt{2}\lambda \underline{\left\vert 0\right\rangle }S_{z}\underline{%
\left\langle 0\right\vert }+2\lambda \sum_{j=2,4,6,\ldots }\underline{%
\left\vert j\right\rangle }S_{z}\underline{\left\langle j\right\vert }.
\notag
\end{eqnarray}%
The subscripts o and e represent contributions associated with the sites
with odd and even parity $\Pi $.\ The structures of $H_{\mathrm{o}}$ and $H_{%
\mathrm{e}}$\ are schematically illustrated in Fig. \ref{fig1}. It indicates
that the invariant space with $k=\pi $\ is split in two unconnected
subspaces. This allow us to investigate the Hamiltonians $H_{\mathrm{o,e}}$\
separately.

\begin{figure*}[tbp]
\includegraphics[ bb=14 94 573 764, width=0.46\textwidth, clip]{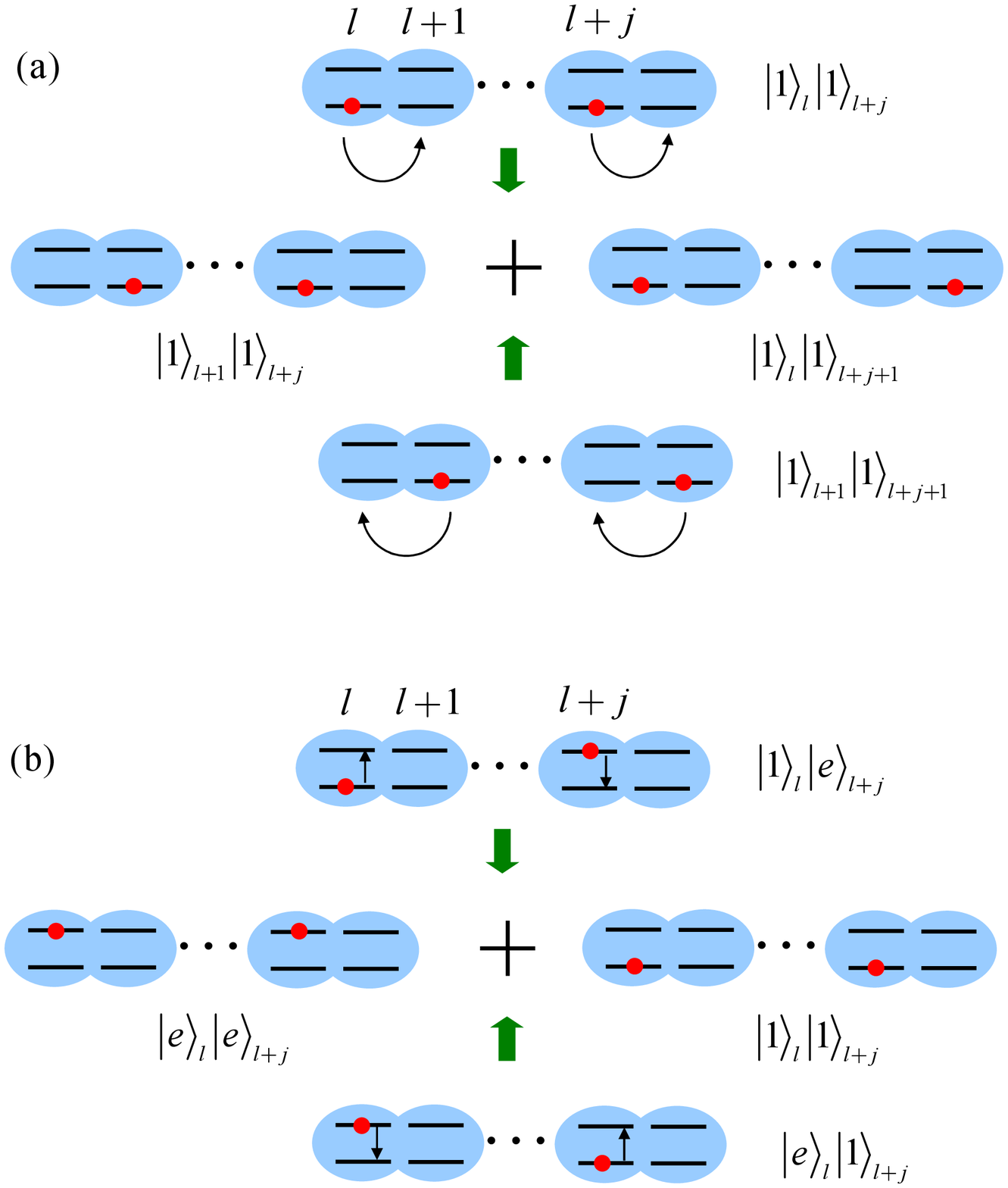} %
\includegraphics[ bb=26 127 566 760, width=0.46\textwidth, clip]{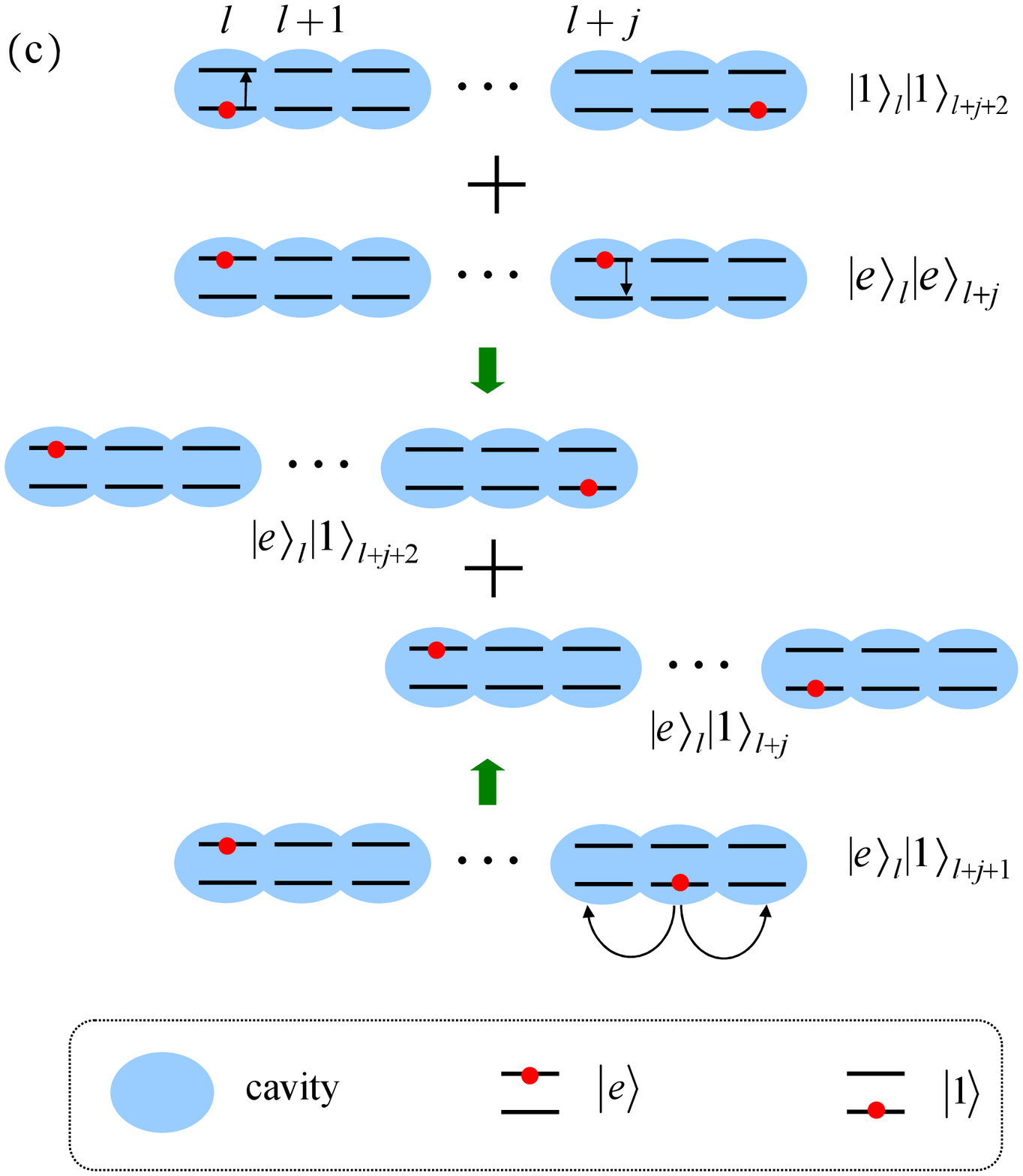}
\caption{(Color online) Schematical illustration for the mechanism of the
formation of bound pair eigenstates. There are three types of destruction
interference processes which result in\ the exact eigenstate $\left\vert
\protect\varphi _{j}\right\rangle $. (a) The Hubbard-type process represeted
in Eq. (\protect\ref{Hubbard type}). (b) The JC-type process represeted in
Eq. (\protect\ref{JC type}). (c) The key process referred as mixed-type in
Eq. (\protect\ref{mixed type}), shows that the cancellation of the
transitions requires the optimal ratio between the parameters $\protect%
\lambda $\ and $\protect\kappa $.}
\label{fig2}
\end{figure*}

\section{Exact bound-pair states}

\label{Exact bound-pair states}

Based on the above analysis, besides states $\left\vert \psi
_{j}\right\rangle $,\ one can also construct a series of bound-pair states
as the form%
\begin{eqnarray}
\left\vert \varphi _{j}\right\rangle &=&\frac{1}{\sqrt{\Omega _{j}}}%
[a_{j}\left( \left\vert j,+\right\rangle -\left\vert j,-\right\rangle \right)
\\
&&+i2\sqrt{2}\left( \lambda /\kappa \right) \left\vert j+1,0\right\rangle
\notag \\
&&-\left( \left\vert j+2,+\right\rangle -\left\vert j+2,-\right\rangle
\right) ],  \notag
\end{eqnarray}%
where the normalization factor $\Omega _{j}$ and amplitudes $a_{j}$
\begin{equation}
\Omega _{j}=2\left( a_{j}\right) ^{2}+8\left( \lambda /\kappa \right) ^{2}+2,
\end{equation}%
and%
\begin{equation}
a_{j}=\left\{
\begin{array}{cc}
2, & j=0 \\
1 & j\geqslant 1%
\end{array}%
\right. .
\end{equation}%
Straightforward derivation shows that%
\begin{eqnarray}
H_{\mathrm{e}}\left\vert \varphi _{j}\right\rangle &=&0,\text{even }j \\
H_{\mathrm{o}}\left\vert \varphi _{j}\right\rangle &=&0,\text{odd }j  \notag
\end{eqnarray}%
i.e., $\left\vert \varphi _{j}\right\rangle $\ is an eigenstate of $H_{%
\mathrm{SO}}$. This is a direct application of the theorem about the bound
state on the graph \cite{Jin}, which states that any eigenstate of a
sub-graph is also the eigenstate of the whole, if the nodes cover all the
joint points. We are interested in the expression of these states\ in the
atom-photon\ basis. It is given by%
\begin{eqnarray}
\left\vert \varphi _{j}\right\rangle &=&\sum_{l}\frac{\left( -1\right) ^{l}}{%
\sqrt{N\Omega _{j}}}[a_{j}\left( \left\vert 1\right\rangle _{l}\left\vert
1\right\rangle _{l+j}+\left\vert e\right\rangle _{l}\left\vert
e\right\rangle _{l+j}\right) \\
&&-2\left( \lambda /\kappa \right) \left( \left\vert 1\right\rangle
_{l}\left\vert e\right\rangle _{l+j+1}-\left\vert e\right\rangle
_{l}\left\vert 1\right\rangle _{l+j+1}\right)  \notag \\
&&+\left( \left\vert 1\right\rangle _{l}\left\vert 1\right\rangle
_{l+j+2}+\left\vert e\right\rangle _{l}\left\vert e\right\rangle
_{l+j+2}\right) ]  \notag
\end{eqnarray}%
for $j\geq 1$, and%
\begin{eqnarray}
\left\vert \varphi _{0}\right\rangle &=&\sum_{l}\frac{\left( -1\right) ^{l}}{%
\sqrt{N\Omega _{0}}}[a_{0}\sqrt{2}\left\vert 2\right\rangle _{l} \\
&&-2\left( \lambda /\kappa \right) \left( \left\vert 1\right\rangle
_{l}\left\vert e\right\rangle _{l+1}-\left\vert e\right\rangle
_{l}\left\vert 1\right\rangle _{l+1}\right)  \notag \\
&&+\left( \left\vert 1\right\rangle _{l}\left\vert 1\right\rangle
_{l+2}+\left\vert e\right\rangle _{l}\left\vert e\right\rangle _{l+2}\right)
].  \notag
\end{eqnarray}%
Alternatively, direct derivation can check our conclusion for the original
Hamiltonian of a lattice atom-photon system in Eq. (\ref{H}) that
\begin{equation}
H\left\vert \varphi _{j}\right\rangle =2\omega _{a}\left\vert \varphi
_{j}\right\rangle .
\end{equation}%
The formation mechanism of these bound-pair eigenstates can be understood as
the result of quantum interference in the following three different types of
processes.

(i) We start with the case of switching off the JC interaction, $\lambda =0$%
. The atoms are decoupled from the cavity array. It is readily to check that%
\begin{equation}
\lbrack \eta _{j},H-\omega _{a}\sum_{l}a_{l}^{\dag }a_{l}]=0,  \label{CR}
\end{equation}%
where the operator $\eta _{j}$\ is defined as%
\begin{equation}
\eta _{j}=\sum_{l}\left( -1\right) ^{l}a_{l}^{\dag }a_{l+j}^{\dagger }.
\label{eta}
\end{equation}%
According to the similiar analysis in Ref. \cite{Yang}, it is found that
state%
\begin{equation}
\left\vert \Psi _{n}\right\rangle =\left( \eta _{j}\right) ^{n}\left\vert
G\right\rangle  \label{n}
\end{equation}%
is an eigenstate of $H$,

\begin{equation}
H\left\vert \Psi _{n}\right\rangle =2n\omega _{a}\left\vert \Psi
_{n}\right\rangle .  \label{S-eq}
\end{equation}%
Furthermore, it is worth to note that even for a Bose Hubbard model, which
involves the on-site interaction,
\begin{equation}
H_{\mathrm{BH}}=-\kappa \sum_{l=1}^{N}\left( a_{l}^{\dag }a_{l+1}+\text{%
\textrm{H.c.}}\right) +\frac{U}{2}a_{l}^{\dag }a_{l}\left( a_{l}^{\dag
}a_{l}-1\right) ,  \label{HB}
\end{equation}%
we still have\
\begin{equation}
\lbrack \eta _{j},H_{\mathrm{BH}}]\left\vert G\right\rangle =0,
\end{equation}%
which leads to the conclusion that $\left\vert \Psi _{1}\right\rangle $ is
an eigenstate of $H_{\mathrm{BH}}$.

The essence of the construction of $\left\vert \Psi _{1}\right\rangle $\ is
due to the destructive interference between the two transitions from states $%
\left\vert 1\right\rangle _{l}\left\vert 1\right\rangle _{l+j}$ and $%
\left\vert 1\right\rangle _{l+1}\left\vert 1\right\rangle _{l+j+1}$%
\begin{equation}
\left.
\begin{array}{c}
H\left\vert 1\right\rangle _{l}\left\vert 1\right\rangle _{l+j} \\
H\left\vert 1\right\rangle _{l+1}\left\vert 1\right\rangle _{l+j+1}%
\end{array}%
\right\} \longrightarrow \left\vert 1\right\rangle _{l+1}\left\vert
1\right\rangle _{l+j}+\left\vert 1\right\rangle _{l}\left\vert
1\right\rangle _{l+j+1},  \label{Hubbard type}
\end{equation}%
which results in%
\begin{equation}
H\left( \left\vert 1\right\rangle _{l}\left\vert 1\right\rangle
_{l+j}-\left\vert 1\right\rangle _{l+1}\left\vert 1\right\rangle
_{l+j+1}\right) \longrightarrow 0.
\end{equation}%
Here the contribution of $H_{\mathrm{0}}$\ is ingored. We refer this as
Hubbard-type process.

(ii) Now we consider the case of switching off the tunneling between
cavities, $\kappa =0$. Each cavity becomes separated from its neighbors. We
have the identity%
\begin{equation}
\left.
\begin{array}{c}
H\left\vert e\right\rangle _{l}\left\vert n-1\right\rangle _{l}\left\vert
n\right\rangle _{l+j} \\
H\left\vert n\right\rangle _{l}\left\vert e\right\rangle _{l+j}\left\vert
n-1\right\rangle _{l+j}%
\end{array}%
\right\} =\lambda n\left\vert n\right\rangle _{l}\left\vert n\right\rangle
_{l+j},  \label{JC type}
\end{equation}%
which results in%
\begin{equation}
H\left( \left\vert e\right\rangle _{l}\left\vert n-1\right\rangle
_{l}\left\vert n\right\rangle _{l+j}-\left\vert n\right\rangle
_{l}\left\vert e\right\rangle _{l+j}\left\vert n-1\right\rangle
_{l+j}\right) =0.
\end{equation}%
This means that there is a destructive interference between the two paths,
which are the atom-photon transitions in the two different cavities $l$ and $%
l+j$.\ It is a pure QED process in a JC model, which is referred as the
JC-type process. It is easy to check that the combination of Hubbard and
JC-type processes result in the formation the eigen state $\left\vert \psi
_{j}\right\rangle $.

(iii) The crucial process that makes state $\left\vert \varphi
_{j}\right\rangle $\ become an eigenstate of the complete Hamiltonian is the
combination of the above two. In this case, the excitation number must be $2$%
. The transitions which result in the destructive interference are%
\begin{eqnarray}
&&\left.
\begin{array}{c}
\left( 1/\lambda \right) \left( \left\vert e\right\rangle _{l}\left\vert
e\right\rangle _{l+j}+\left\vert 1\right\rangle _{l}\left\vert
1\right\rangle _{l+j+2}\right)  \\
\left( -1/\kappa \right) \left\vert e\right\rangle _{l}\left\vert
1\right\rangle _{l+j+1}%
\end{array}%
\right\}   \label{mixed type} \\
&\rightarrow &\left( \left\vert e\right\rangle _{l}\left\vert 1\right\rangle
_{l+j}+\left\vert e\right\rangle _{l}\left\vert 1\right\rangle
_{l+j+2}\right) .  \notag
\end{eqnarray}%
We note that the cancellation occurs only if the amplitudes of the two
components $\left\vert e\right\rangle _{l}\left\vert e\right\rangle
_{l+j}+\left\vert 1\right\rangle _{l}\left\vert 1\right\rangle _{l+j+2}$\
and $\left\vert e\right\rangle _{l}\left\vert 1\right\rangle _{l+j+1}$\ are
properly assigned. We refer this to the mixed-type process.\ In Fig. \ref%
{fig2}, three processes for the formation mechanism of the bound pair state
is schematically illustrated.

\section{Long-range entanglement}

\label{Long-range entanglement}

We now study the feature of the obtained eigenstates. Apparently, the pair
state $\left\vert \psi _{j}\right\rangle $ and $\left\vert \varphi
_{j}\right\rangle $ are entangled states. In the strong coupling limit $%
\lambda \gg \kappa $, we have%
\begin{equation}
\left\vert \varphi _{j}\right\rangle \approx \sum_{l}\frac{\left( -1\right)
^{l}}{\sqrt{2N}}\left( \left\vert 1\right\rangle _{l}\left\vert
e\right\rangle _{l+j+1}-\left\vert e\right\rangle _{l}\left\vert
1\right\rangle _{l+j+1}\right) ,
\end{equation}%
which is the superposition of entangled states between two cavities at
distance $j+1$. States
\begin{equation}
(\left\vert 1\right\rangle _{l}\left\vert e\right\rangle _{l+j+1}-\left\vert
e\right\rangle _{l}\left\vert 1\right\rangle _{l+j+1})/\sqrt{2}
\end{equation}%
in $\left\vert \varphi _{j}\right\rangle $\ and
\begin{equation}
(\left\vert e\right\rangle _{l}\left\vert e\right\rangle _{l+j}-\left\vert
1\right\rangle _{l}\left\vert 1\right\rangle _{l+j})/\sqrt{2}
\end{equation}%
in $\left\vert \psi _{j}\right\rangle $\ are both maximally entangled states
of $l$-th and $\left( l+j\right) $-th\ (or $\left( l+j+1\right) $-th)
cavities for the two modes, excited cavity field and excited atom modes. To
demonstrate this concept in a precise manner, we introduce lower branch and
upper branch exciton-polariton states,%
\begin{eqnarray}
\left\vert \downarrow \right\rangle _{l} &=&\frac{1}{\sqrt{2}}(i\left\vert
1\right\rangle _{l}-\left\vert e\right\rangle _{l}),  \label{lower} \\
\left\vert \uparrow \right\rangle _{l} &=&\frac{1}{\sqrt{2}}(i\left\vert
1\right\rangle _{l}+\left\vert e\right\rangle _{l}),  \label{upper}
\end{eqnarray}%
the superposition of which yields a polariton qubit state at cavity $l$.
With $\left\vert \downarrow \right\rangle _{l}$\ and $\left\vert \uparrow
\right\rangle _{l}$\ being basis, it is given that

\begin{eqnarray}
\left\vert \varphi _{j}\right\rangle &\sim &\frac{1}{\sqrt{2}}(\left\vert
\uparrow \right\rangle _{l}\left\vert \uparrow \right\rangle
_{l+j+1}-\left\vert \downarrow \right\rangle _{l}\left\vert \downarrow
\right\rangle _{l+j+1}), \\
\left\vert \psi _{j}\right\rangle &\sim &\frac{1}{\sqrt{2}}(\left\vert
\uparrow \right\rangle _{l}\left\vert \uparrow \right\rangle
_{l+j}+\left\vert \downarrow \right\rangle _{l}\left\vert \downarrow
\right\rangle _{l+j}),
\end{eqnarray}%
which are standard Bell states. We note that the entanglement does not
decrease as the distance $j$\ increases. The entanglement is one of the
great importance for the new field of quantum information theory. Polaritons
\cite{Khitrova} as quasiparticles of light and matter, are the most
promising solution for the interface between electronic and photonic qubit
states.

However, we would like to point out that two atoms for state $\left\vert
\psi _{j}\right\rangle $ (or $\left\vert \varphi _{j}\right\rangle $), in $l$%
-th and $\left( l+j\right) $-th\ (or $\left( l+j+1\right) $-th) cavities, do
not entangle with each other due to the following reason. The atomic
entanglement can be characterized by concurrence \cite{Huo}\textit{. }The
reduced density matrix for two atoms in $l$-th and $\left( l+j+1\right) $-th
cavities is
\begin{equation}
\rho ^{(l,l+j+1)}=\mathrm{Tr}_{p}\mathrm{Tr}_{(l,l+j+1)}\left( \left\vert
\varphi _{j}\right\rangle \left\langle \varphi _{j}\right\vert \right) ,
\label{density matrix}
\end{equation}%
where $\mathrm{Tr}_{p}$ denotes the trace over all photon variables and $%
\mathrm{Tr}_{(l,l+j+1)}$\ denotes the trace over all atomic variables except
for $l$-th and $\left( l+j+1\right) $-th atoms. It has been shown in Refs.
\cite{Huo} that the formula for the concurrence of two quasi-spin in a
hybrid system is the same as that for pure spin-$1/2$ system \cite{Wootters,
Wang}. Then the concurrence $C_{ll^{\prime }}$ shared between two atoms $l$
and $l^{\prime }$ is obtained as

\begin{equation}
C_{ll^{\prime }}=2\max (0,\left\vert z_{ll^{\prime }}\right\vert -\sqrt{%
u_{ll^{\prime }}^{+}u_{ll^{\prime }}^{-}}).  \label{c2}
\end{equation}%
in terms of the quantum correlations
\begin{eqnarray}
z_{ll^{\prime }} &=&\left\langle \varphi _{j}\right\vert \sigma
_{l}^{+}\sigma _{l^{\prime }}^{-}\left\vert \varphi _{j}\right\rangle , \\
u_{ll^{\prime }}^{\pm } &=&\frac{1}{4}\left\langle \varphi _{j}\right\vert
\left( 1\pm \sigma _{l}^{z}\right) \left( 1\pm \sigma _{l^{\prime
}}^{z}\right) \left\vert \varphi _{j}\right\rangle ,
\end{eqnarray}%
where $\sigma _{i}^{+}=\left( \sigma _{i}^{-}\right) ^{\dag }=\left\vert
e\right\rangle _{i}\left\langle g\right\vert $. It is a straightforward
calculation to show that $C_{ll^{\prime }}$\ is always zero for both states $%
\left\vert \varphi _{j}\right\rangle $\ and $\left\vert \psi
_{j}\right\rangle $.

\section{Summary}

\label{Summary}

In summary, we have established the link between the two-excitation JCH
model and the single-particle $4$-leg ladder with an effective flux, which
is shown to be equivalent to a chain system of spin-$1$\ particle with
spin-orbit coupling. It also introduces a mechanism to construct a series of
bound-pair eigenstates, which display long-range polaritonic entanglement.
This finding reveals that the hybrid system can offer rich features and
useful functionality, which will motivate further investigation.

\acknowledgments We acknowledge the support of the National Basic Research
Program (973 Program) of China under Grant No. 2012CB921900 and CNSF (Grant
No. 11374163).

\end{document}